\documentstyle[12pt]{article}
\begin{document}

\begin{flushright}
OSU-TA-97-3 \\
February 1997
\end{flushright}

\vspace{1in}
\begin{center}
{\Large{\bf Calculation of the emergent spectrum  and observation of
primordial black holes.}}\\

\vspace{.4in}

{\bf Andrew F. Heckler} \\
{\em  Astrophysics and Cosmology Group,
Ohio State University,}\\
{\em Columbus, Ohio USA}\\{\em email: heckler@mps.ohio-state.edu}
 
\vspace{.2in}
\begin{abstract}

We calculate the emergent spectrum of microscopic black holes, which
emit copious amounts of thermal ``Hawking'' radiation, taking into
account the proposition that (contrary to previous models) emitted
quarks and gluons do not directly fragment into hadrons, but rather
interact and form a photosphere and decrease in energy before
fragmenting. The resulting spectrum emits copious amount of photons at
energies around 100MeV. We find that the limit on the average
universal density of black holes is not significantly affected by the
photosphere. However we also find that gamma ray satellites such as
EGRET and GLAST are well suited to look for nearby black holes out to
a distance on the order of 0.3 parsecs, and conclude that if black
holes are clustered locally as much as luminous matter, they
may be directly detectable.

\end{abstract}
\end{center}

\vspace{.6in}
submitted to {\em Physical Review Letters}
\renewcommand{\thesection}{\Roman{section}} 

\newpage

\section{Introduction}

Since Hawking \cite{Hawking74,Hawking75a} first proposed that a black
hole emits thermal radiation with an emission rate inversely
proportional to its mass, there have been several calculations of the
emergent spectrum in order to verify or at very least constrain the
presence of the smallest, hence most luminous ones. Nominally, aside
from particle spin effects and gravitational backscattering effects
close to the black hole, calculated by Page \cite{Page76a}, one would
expect the the emergent spectrum to be thermal, since Hawking showed
that the black hole can be thought of as a black body with a
temperature $T = (8\pi G M)^{-1}$, where $M$ is the mass of the black
hole, $G$ is Newton's gravitational constant, and we set the Boltzmann
constant, $k =1$ (along with $c=1$ and $\hbar =1$).

However, when a detailed model of the physics of the emitted particles
in considered, the emergent spectrum becomes more complicated. For
example, MacGibbon and Weber \cite{MW} and Halzen et al. \cite{HZ} have
considered black holes with temperatures greater than the
characteristic QCD energy scale $\Lambda \sim 200$MeV, where the black
hole begins to emit quarks and gluons. They propose that the emitted
quarks and gluons fragment into hadrons, which further decay into
photons, electrons, neutrinos etc., and they convolve a jet code with
the Hawking thermal spectrum to determine the emergent spectrum. For
black hole temperatures above $\Lambda$, the QCD degrees of freedom
dominate in the standard model and the ultimate products of quark and
gluon fragmentation will dominate the spectrum, thus understanding the
physics of the quarks and gluons is important for determining the
spectrum.

In this letter, we reconsider an important initial assumption made by
MacGibbon and Weber, namely that the quarks and gluons emitted from
the black hole directly fragment into hadrons. Rather, since the
density of emitted particles around the black hole can be very high
(much higher than nuclear density) for $T>\Lambda$, we propose that
the quarks and gluons propagate through this dense plasma and lose
energy via QCD bremsstrahlung and pair production interactions until
the density of the outward-propagating plasma becomes low enough that
the quarks finally fragment into hadrons. One can then calculate an
emergent spectrum by convolving this collection of lower energy quarks
with jet codes or fragmentation functions, but this spectrum will be
very different than the one obtained using fragmentation of quarks
coming directly from the black hole. In addition to determining the
emergent spectrum of the black hole including the effect of
interactions, we will also consider several observational consequences
of the this emergent spectrum, using the EGRET and GLAST satellites as
exemplary detectors.
%This effect is similar to the
%effect of the quark gluon plasma on jet production in heavy ion
%collisions. In this case, higher energy jets are not as abundant as in
%the simpler proton/anti-proton collisions because the high energy
%quarks interact with the nuclear matter before fragmenting
%into hadrons.

Since it is reasonable to assume that fundamental modes will be
present in a thermal bath, let us assume that once above a temperature
$T>\Lambda$, a black hole emits individual quarks and gluons. Hawking
showed that the emission rate spectrum for each particle degree of
freedom is $d{\dot N}/dE = \Gamma_{s}/2\pi (\exp(E/T)-
(-1)^{s})^{-1}$, for particles of energy $E$, where $\Gamma_{s}$ is
the absorption coefficient which in general depends upon the spin of
the particle $s$, $M$, and $E$ \cite{Page76a}, and we have assumed the
black hole to be uncharged and non-rotating. In the relativistic limit
$T \gg m$ the total rate at which particles are {\em directly} emitted
from the black hole can be expressed in the form ${\dot N}_{\rm tot} =
(10^{-2}\eta)T$, where $\eta$ is of order of the number of emitted
relativistic particles at temperature $T$ and can be calculated
numerically \cite{Page76a,MW}. Since this is the flux of particles
crossing the Schwartzschild radius $r_{s}\equiv (4\pi T)^{-1}$, the
density $n(r)$ of emitted particles at a radius $r$ from the black
hole is then $n = {\dot N}/(4\pi r^{2})$. Expressing this in a more
illuminating form, we obtain
\begin{equation}\label{density}
n(r) = \left(\frac{4\pi\eta}{100}\frac{r_{s}^{2}}{r^{2}}\right)T^{3}.
\end{equation}
For the QCD, $\eta \sim 20$, thus the average
particle separation, defined as $d \equiv n^{-1/3}$ is then $d(r)
\simeq T^{-1} (r_s/r)^{2/3}$. 
%One can also estimate the average
%difference in momentum $\bar{\Delta p}$ between the outward
%propagating particles, which are emitted in random directions, by
%noting that the average angle subtended between nearest particles is $d/r$ and
%since the average momentum of the emitted particles is of order $T$,
%one finds $\bar{\Delta p} \sim Td/r \sim T(r_s/r)^{1/3}$.

QCD is an asymptotically free theory of interactions between quarks
and gluons. In general, when particles scatter, the momentum exchanged
must be at least of order their inverse separation $d^{-1}$. As a
consequence, at distances smaller than $\Lambda^{-1}$ the QCD
interaction is perturbative, while at larger distances, the coupling
becomes so large that vacuum polarization and fragmentation is
dominant. In particular, vacuum fragmentation of quarks and gluons
will occur when they are separated by a distance greater than
$\Lambda^{-1}$. We see from the above formulas that once $T>\Lambda$,
the quarks and gluons around the black hole will not immediately
vacuum fragment into hadrons because they are closely spaced in a kind
of plasma. Rather, they will propagate for some distance in the dense
quark-gluon plasma until the plasma becomes tenuous enough that vacuum
fragmentation will occur. This is important because as the quarks and gluons
propagate away from the black hole in the dense plasma, they will
interact with each other via bremsstrahlung and pair production and
decrease in energy.

This can be seen by following the arguments of Heckler \cite{photo},
who has shown that QED and QCD bremsstrahlung and pair production
interactions become important among particles emitted from a black
hole above some critical temperature. The essential argument stems
from the fact that the relativistic bremsstrahlung cross section is
independent of energy, and since the density of particles around the
black hole increases with temperature (eq.~\ref{density}), there is a
critical temperature for which the optical depth of an emitted
particle becomes unity. At this point particles begin to scatter
copiously, and a kind of photosphere forms around the black hole. The
photosphere is a kind of fireball in the sense that the nearly
thermalized plasma propagates outward, decreasing in temperature until
i) the electrons and positrons annihilate in the QED case or ii) the
quarks and gluons fragment or some kind of QCD phase transition occurs
in the QCD case.

To simplify matters, we will assume that the plasma cools to the
temperature $\Lambda$, at which point hadronization occurs. The
emergent photon spectrum is a convolution of the quark and gluon
spectrum with the pion fragmentation function \cite{MW,HZ} and the
Lorentz-transformed neutral pion decay into photons
\begin{equation}
\frac{d\dot{N}_{\gamma}}{dE_{\gamma}}
=\int_{m_{\pi}}^{
(4E_{\gamma}^{2}+m_{\pi}^2)/4E_{\gamma}}{\frac{d\dot{N}_{\pi}}{dE_{\pi}}\frac{dg_{\pi\gamma}(E_{\pi})}{dE_{\gamma}}dE_{\pi}}
\end{equation}
where $dg_{\pi\gamma}/dE = 2/(\gamma m_{\pi}\beta)$ is the number of
photons of energy $E$ created by an pion decaying isotropically in its
rest frame, $\gamma = (1-\beta^2)^{-1}$, and $\beta$ is the velocity
of the pion (note $E_{\gamma}$ is Doppler shifted,
$m_{\pi}/(2\gamma(1+ \beta))>E_{\gamma}>m_{\pi}/(2\gamma(1-
\beta))$) . The pion spectrum is \cite{MW}
\begin{equation}
\frac{d\dot{N}_{\pi}}{dE_{\pi}} =
\sum_{j}\int_{Q=E_{\pi}}^{Q=\infty}{\frac{d\dot{N}_{j}(Q,T_0)}{dQ}\frac{dg_{j\pi}(Q,E_{\pi})}{dE_{\pi}}dQ}
\end{equation}
where $d\dot{N}_{j}/dQ$ is the flux spectrum of the quark or gluon $j$
of energy $Q$ at the time of fragmentation, which is on the outer edge
of the photosphere where the plasma is at a temperature of
$T_{0}=\Lambda$, and $dg_{j\pi}/dE_{\pi}$ is the relative number of
pions with energy $E_{\pi}$ produced by $j$ \cite{MW}. We will use
$dg/dE_{\pi} = (15/16)z^{-3/2}(1-z)^{2}$, where $z=E_{\pi}/Q$
\cite{HZ}. We approximate the quark and gluon spectrum in the observer
frame by  boosting a thermal spectrum at temperature
$T_{0}$ with the lorentz gamma factor $\gamma_{p}$ of the outer edge
of the outward moving photosphere
\begin{equation}
\frac{d\dot{N}_{j}}{dQ} =
\sigma_{j}\frac{\gamma_{p}^{2}r_{p}^{2}Q^{2}}{2\pi^{2}}\int_{\cos{\theta}=0}^{1}{\frac{(1-\beta\cos{\theta})\cos{\theta}}{\exp{(\gamma_{p}\omega
(1-\beta\cos{\theta})/T_{0})}\pm 1}d\Omega}
\end{equation}
%where $\sigma_{j}= 6$ for quarks (12 quarks and anti-quarks total) and
%$\sigma_{j} = 2$ for gluons (8 gluons total),
where $\sigma_{j}$ is the number of internal degrees of freedom of
particle $j$, the sign in the denominator is for fermions or bosons,
and we have we integrated over the surface of the photosphere with
radius $r_{p}$. Using ref. \cite{photo}, we approximate $\gamma_{p} \simeq
(T/\Lambda)^{1/2}$ and $r_{p} = \gamma_{p}/\Lambda$. An accurate
calculation of the spectrum would require using a boltzmann equation
to determine the exact spectrum of the nearly thermal quarks and
gluons, and a jet fragmentation code for the final decay into pions
and photons. We expect the approximations used will be correct within
a factor of order unity.

In Figure~1 we show the photon spectrum calculated using the above
formulas both including and excluding the QCD photosphere. Without the
photosphere, the quark and gluon spectrum is simply a blackbody
spectrum (with spin and finite size effects) at the temperature
$T$. We see that the main difference between these two spectra is that
the photosphere spectrum has many more lower energy photons. This is
physically due to the fact that the photosphere processes high energy
quarks and gluon into many lower energy ones, and these eventually
fragment and decay into lower energy photons. We also plot the
spectrum of photons emitted directly from the black hole which peak at
approximately $5T$. The peak height is several orders of magnitude
lower than the peak from quarks and gluons, mostly due to the QCD
degrees of freedom which can decay into photons is large \cite{MW}. The
total number of photons emitted from the QCD photosphere, which peaks
at energies of about 100MeV, is
\begin{equation}\label{Ndot}
\dot{N}_{\gamma}\approx 2\times 10^{24}\left(\frac{T}{{\rm
GeV}}\right)^{2}{\rm sec}^{-1}
\end{equation}
which scales as $T^2$, which is a stronger function of black hole
temperature than the $T$ dependence of black body particle emission,
or the $T^{3/2}$ dependence obtained in the fragmentation model of
MacGibbon and Weber. Note that once the black hole is above the QED
critical temperature of about 45GeV, the direct photons will also be
processed through a QED photosphere and be degraded to low energies
\cite{photo}. We have included the QED photosphere spectrum in figure~1,
which peaks between 1 and 10 MeV and produces $\dot{N}_\gamma \approx
5\times 10^{28}(T/100{\rm GeV})^{3/2} {\rm s}^{-1}$. The extra power
of $T^{1/2}$ in the QCD case comes from the multiplicity of
fragmentation of quarks into pions.

Let us examine several observational consequences of the QCD
photosphere.  The most important consequence involves the search for
individual nearby primordial black holes. When developing a strategy
and interpreting the results of a direct search for expiring black
holes, one must consider the density, emergent spectrum and lifetime
of the black holes, all of which are a function of black hole mass,
and the detector sensitivity and background, which are a both a
function of photon energy. First, one can determine the optimum energy
range to search for these black holes by considering the background :
since the observed gamma ray background scales approximately as
$E^{-2.4}$ in the range 1MeV to 10GeV, one can show using the example
of the spectrum of figure~1 that for black holes with $T<$ 10TeV
(higher $T$ black holes have lifetimes shorter than 1s, and emit
negligible amounts of radiation), the optimum signal to background
lies in the range of 1 to 10GeV.

Next, one can determine the optimum black hole mass to which the
detector is sensitive. Naturally, if $I_{BH}$ is the photon emission
rate of the black hole, and $\ell$ is the limiting flux to which the
detector is sensitive, then $\ell <I_{BH}/4\pi d^2$, where $d$ is the
distance to the black hole. We know that, $I_{BH} \propto M^{-2}$,
however, the lifetime of the black hole $\tau \propto M^3$. One must,
therefore, also require that the total lifetime integrated number of
photons incident on the detector is (at {\em least}) greater than
unity. To be more realistic, let us require the observed amount of photons
$N_{\gamma}>10$. Thus, $(I_{BH}/4\pi d^2)A \tau > 10$, where $A$ is
the area of the detector. When both conditions are met, one finds
$\tau >10/(\ell A)$. The optimal observing conditions thus occur for
black holes which just meet this criterion. By using eq.~\ref{Ndot}
and noting that $\tau \approx M^3/3\alpha$ (see e.g. \cite{HZ}), we
can roughly estimate the maximum observing distance $d_{\rm max} \sim
0.2\,{\rm pc}\,(A/2000{\rm cm}^{2})^{1/3}(10^{-9}{\rm cm}^{-2}{\rm
s}^{-1}/\ell)^{1/6}$.

Let us consider two satellites: EGRET \cite{egret}, which has already
accumulated several years of data, and GLAST \cite{glast}, which is
still in its planning stages. In figure~2 we show the maximum distance
each satellite can observe a small black hole, and determine that for
EGRET (GLAST), $d_{\rm max} \simeq 0.11\,\, (0.31)\,{\rm pc}$.

Once one has found a limit on the maximum distance one can observe
these black holes, one can refashion this limit into other useful
limits. MacGibbon and Carr \cite{MC}, and Halzen et al. \cite{HZ} use
the Page-Hawking limit \cite{PH}, which is discussed below, to find a
generic limit on the local density of black holes below a mass $M$ to
be $n_{bh}< {\cal N}(\zeta /3)(M/M_{*})^3 {\rm pc}^{-3}$, where
$M_{*}\approx 5\times 10^{14}$g is the mass of a black hole which has
a life of the age of the universe, ${\cal N}\approx 10^{-4}{\rm
pc}^{-3}$ is Page-Hawking limit on the average density of $M<M_{*}$
black holes, and $\zeta$ is the local density enhancement of black
holes compared to the universal average. This general scaling solution
is valid up to masses $M_{*}$ and is related to the fact that the
black hole lifetime is proportional to $M^3$. We can make a similar
limit on the value of $\zeta {\cal N}$ by assuming that if EGRET and
GLAST find no black holes after a time $\tau =(\ell A)^{-1}$, then
this will optimally constrain the density black holes with masses
$M_{0}$ whose lifetime is $\tau$. However, there is a subtle
beneficial effect: after observing for a time $t>\tau$, larger mass
black holes, which have higher number densities, will have decayed to
mass $M_{0}$. Since $dM/dt = -\alpha/M^2$ (see e.g. \cite{HZ}), one
can begin to constrain black holes with mass $M^3 = 3\alpha \delta t +
M_{0}^3$, where $\delta t = t-\tau$. If the constraint on the density
of $M_{0}$ black holes is $n_{0}<(\Omega d_{0}^3/3)^{-1}$, where
$\Omega$ is the solid angle covered by the detector, then we obtain
the limit
\begin{equation}\label{zetaN}
\zeta {\cal N} < \frac{9}{\Omega
d^{3}}\left(\frac{M_{*}^{3}}{M_{0}^{3}+3\alpha \delta t}\right)
\approx \frac{1.5\times 10^{10}}{\Omega d^3}\left(\frac{{\rm
yr}}{\delta t}\right).
\end{equation}
as long as $t> \tau$. Simply put, observing for a longer time allows
one to observe larger, more densely populated black holes which decay
to the optimum observing mass $M_{0}$, and this allows for better limits on
$\zeta {\cal N}$. With four years of observation and assuming no black
holes are found with EGRET or GLAST, one can place the limits
\begin{eqnarray}
\zeta {\cal N}&<& 5.7\times 10^{12}\, {\rm pc}^{-3}\,\,{\rm (EGRET)}\nonumber\\
\zeta {\cal N}&<& 8.5\times 10^{10}\, {\rm pc}^{-3} \,\,{\rm (GLAST)}
\end{eqnarray}
Notice that this limit scales as
$\sqrt{\ell}/A$. As pointed out by Halzen et al. \cite{HZ}, luminous
matter clusters locally by a factor on the order of $10^{7}$, thus
since ${\cal N} < 10^{4}$, EGRET may be able to see nearby primordial
black holes. One can also place a limit on the local rate $R$ at
which black holes expire per unit volume.  Roughly, $R \approx
n(M)/\tau(M)\approx \zeta {\cal N} \alpha/M_{*}^{3}$. For EGRET
(GLAST) one finds $R< 1100\,\, (17)\, {\rm pc}^{-3}{\rm yr}^{-1}$.

Comparison of these results to previous results is difficult because
non-standard particle models were often used. For example, there have
been several limits placed on $R$ based on the Hagedorn model \cite{Hag},
which presumes that the number of degrees of freedom increase
exponentially with temperature, a fact that has no experimental
support for energies up to about 1TeV.
It is important to note, however, that Alexandreas
et al. \cite{Alex}  use the standard model of MacGibbon and Webber,
and conclude from data of a TeV air shower array that $R< 8.5\times
10^5 {\rm pc}^{-3}{\rm yr}^{-1}$. However, as illustrated in figure~3,
the photosphere dramatically alters the lifetime integrated spectrum,
and the expected flux of photons above 1TeV is about four orders of
magnitude lower than the MacGibbon and Webber result, which translates
into a limit on $R$ six orders of magnitude weaker (higher). Note
especially that when the photosphere is included, the lifetime
integrated spectrum decreases as $E^{-4}$ instead of $E^{-3}$. In fact
the total number of photons above energy $E_{d}$ produced by a black
hole of initial temperature $T$ is $N_{\gamma} \approx 2\times 10^{34}
({\rm GeV}/E_{d})^{3}$, which is valid for energies $E_{d}>(T/{\rm
GeV})^{1/2}$GeV. This is to be compared to the result of Halzen et al,
which finds $N_{\gamma}\propto E_{d}^{-2}$. This lower flux at high
energies renders any search for primordial black holes with TeV air
shower arrays impractical.

It is interesting to note that Wright \cite{wright} points out that
the anisotropic component of the gamma ray background may be explained
by the presence of primordial black holes clustered in the halo of our
galaxy. If black holes are responsible for the anisotropy, then he
finds $\zeta {\cal N} = (2-12)/h\times 10^{9}\,{\rm pc}^{-3}$, where
$h$ is the Hubble parameter, and this is only about an order of
magnitude lower than the estimated detection limits of the GLAST
project.

Another consequence involves the Page-Hawking limit, which
constrains (or possibly measures) the density of primordial black
holes by comparing their expected contribution to the gamma ray
background, for a given black hole density, to the actual observed
background. MacGibbon and Carr \cite{MC} , and Halzen et al. \cite{HZ},
have studied this approach in detail, using a model of direct
fragmentation of quarks and gluons (i.e. no photosphere), and they
conclude that black holes with mass less than about $10^{16}$g cannot
contribute more than a fraction $\Omega_{bh}< 10^{-8}$ of the critical
density of the universe.

We have performed the same calculation including the effects of the
QCD photosphere (see figure~3) and verified, as suggested by Heckler
\cite{photo}, that the QCD photosphere does not significantly change
(i.e. less than a few percent) the limit on $\Omega_{bh}$ found by
Halzen et al. \cite{HZ}. The effect is small because the QCD
photosphere becomes important only at temperatures above $\Lambda$,
and black holes with these temperatures do not significantly
contribute to the total number of photons at 100MeV. There is one
important difference in the spectrum: above energies of about 300MeV
the spectrum including the photosphere is proportional to $E^{-4}$,
whereas the results of Halzen et al. show a $E^{-3}$ dependence. The
steeper slope is due to the degrading of high energy quarks into
lower energies as they are processed through the photosphere.

As a final note, since the QCD photosphere will emit charged pions as
well as neutral ones, there will also be a flux of neutrinos,
electrons, and positrons up to several orders of magnitude larger than
previously assumed, with spectra very similar to the photon spectrum
in Figure~1. This will make constraints on (or possibilities of
detection of) high energy neutrinos \cite{HKZ}, and positrons \cite{MC}
from black holes much more important.

This work was supported by DOE grant DE-FG02-91ER40690 at THE Ohio
State University.

\begin{figure}[ht] 
\input{epsf}
\centerline{\epsfxsize 6.5in \epsfbox{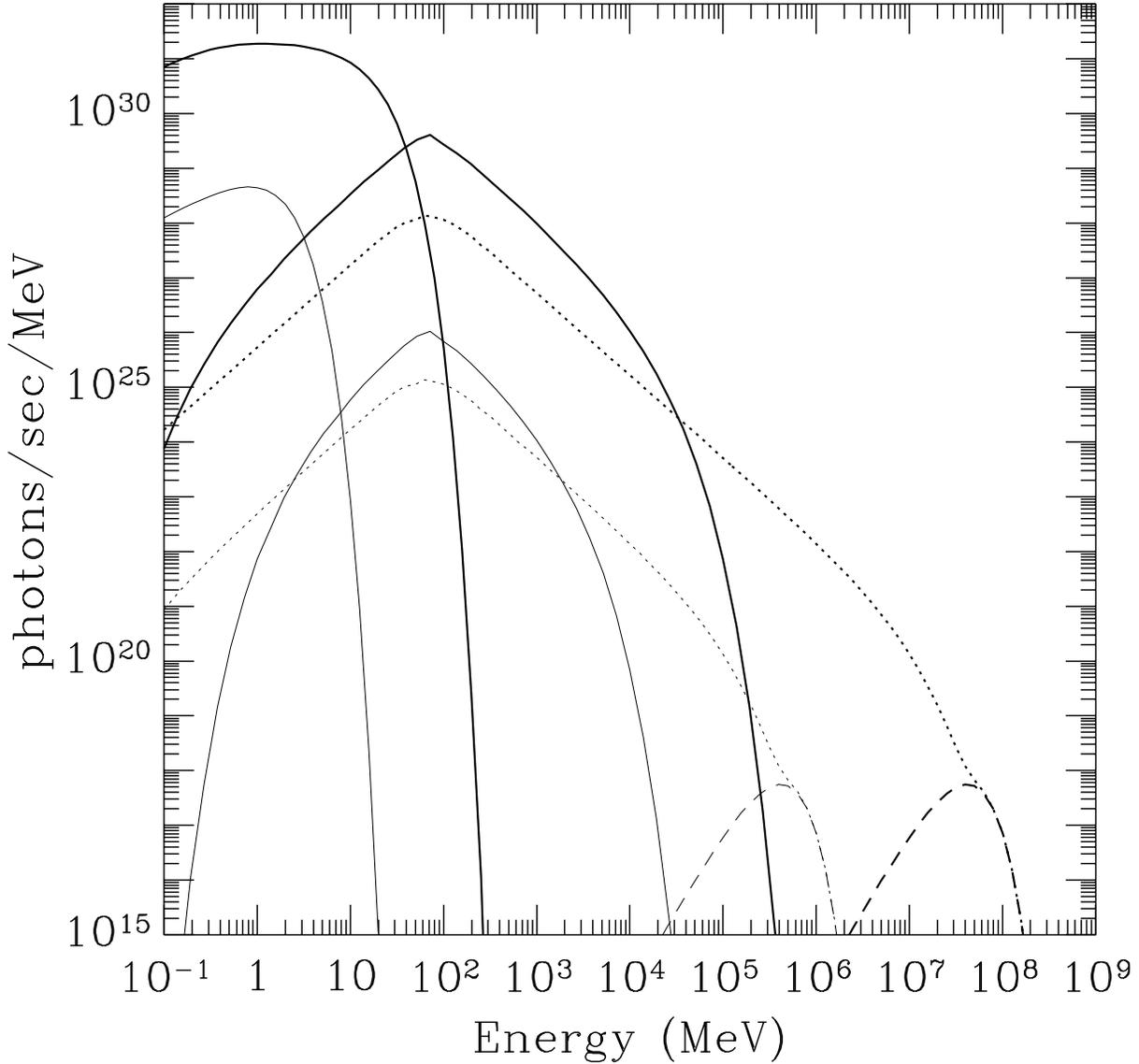}}
\caption{Instantaneous emergent spectra. Thick lines are for a
$M=10^{9}$g $(T = 10$TeV) black hole. Thin lines for $M=10^{11}$g $(T
= 100$GeV). The solid lines are spectra which include photosphere: the
ones peaking at about 100MeV are the emergent spectra of QCD
photosphere, the ones peaking at about 1MeV are for the QED
photosphere. For comparison, we plot the dotted lines which are the
direct fragmentation results of MacGibbon and Webber, and the dashed
lines which are the direct photon emission spectra. The actual full spectrum
is the addition of the two solid lines.
\label{fig:inst}}
\end{figure}

\begin{figure}[ht] 
\input{epsf}
\centerline{\epsfxsize 6.5in \epsfbox{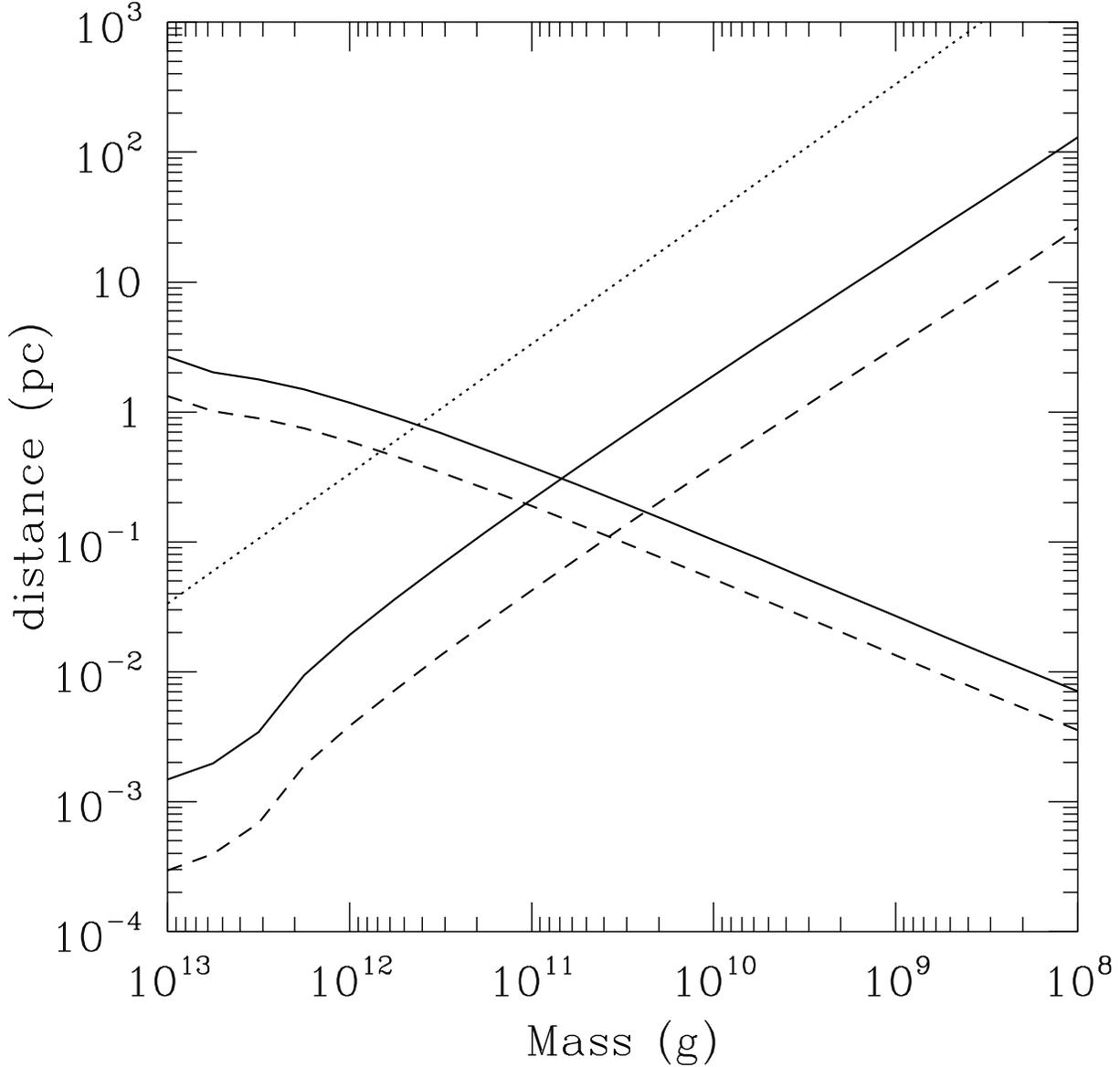}}
\caption{Observational distance limits. The dotted line is the minimum
average distance to the nearest black hole of mass $M$, using the
Page-Hawking constraint and assuming a clustering
factor $\zeta = 10^{6}$. The solid (dashed) lines pertain to GLAST
(EGRET). Lines sloping upward show the maximum distance a black hole
can be detected for the GLAST (EGRET) point source sensitivity, and
the line sloping downward shows the maximum distance a black hole can
be and still cast 10 photons on the detector, integrated over
the lifetime of the hole. As the observation time of the detector
increases, the intersection of the lines effectively moves to the left
(see eq.~\ref{zetaN}), possibly intersecting with the Page-Hawking constraint.
\label{fig:dmax}}
\end{figure}

\begin{figure}[ht] 
\input{epsf}
\centerline{\epsfxsize 6.5in \epsfbox{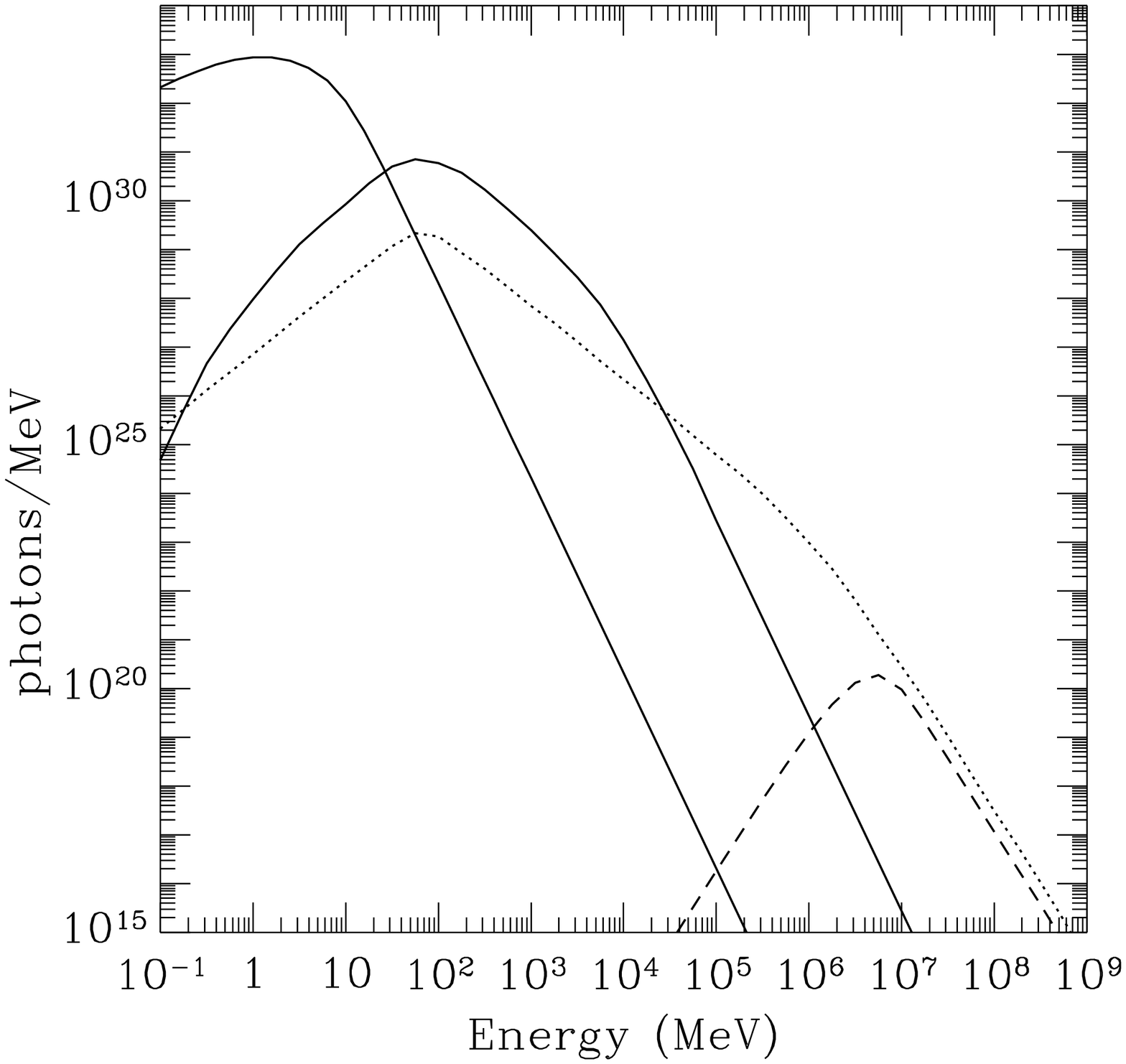}}
\caption{ Lifetime integrated emergent spectra for a $M=10^{10}$g $(T=
1$ TeV) black hole, which has a lifetime of about 400s. Notation is
the same as in Figure~1. Notice that the slope of the photosphere
spectra (solid line) runs as $E^{-4}$, whereas for the direct
fragmentation picture dotted line) the slope runs as $E^{-3}$.
\label{fig:ob}}
\end{figure}


\begin{thebibliography}{99}

\bibitem{Hawking74} S.W. Hawking, Nature {\bf 248}, 30 (1974).

\bibitem{Hawking75a} S.W. Hawking, Commun. Math. Phys. {\bf 43}, 199 (1975).

\bibitem{Page76a} D.N. Page, Phys. Rev. {\bf D13}, 198 (1976).

\bibitem{MW} J.H. MacGibbon and B.R. Webber, Phys. Rev. {\bf D41},
3052 (1990).

\bibitem{HZ} F. Halzen, E. Zas, J.H. MacGibbon \& T.C. Weekes, Nat. {\bf 353}, 807 (1991).

\bibitem{photo} A.F. Heckler, Phys.Rev. {\bf D55}, 480 (1997).

\bibitem{egret} G. Kanbach et al., Space Sci. Rev. {\bf 49}, 69 (1988).

\bibitem{glast} P.F. Michelson, Proc. SPIE {\bf 2806}, 31 (1996).

\bibitem{MC} J.H. MacGibbon and B.J. Carr, Astrophys. J. {\bf 371}, 447 (1991).

\bibitem{PH} D.N. Page and S.W. Hawking, Astrophys. J. {\bf 206}, 1
(1976).

\bibitem{Hag} R. Hagedorn, Nuovo Cimento {\bf 56}, Ser. A, 1027 (1968);
R. Hagedorn, A\&A {\bf 5}, 184 (1970).

%\bibitem{rees} M.J. Rees, Nat. {\bf 266}, 333 (1977).

\bibitem{Alex} D.E. Alexandreas {\em et al.}, Phys. Rev. Lett. {\bf 71}, 2524 (1993).

\bibitem{wright} E.L. Wright, Ap.J. {\bf 459}, 487 (1996).

\bibitem{HKZ} F. Halzen, B. Keszthelyi, \& E. Zas, Phys.Rev. {\bf
D52},  3239 (1995).

\end{thebibliography}
\end{document}